\documentclass[aps,prb,twocolumn]{revtex4}
\usepackage{graphicx}
\newcommand\beq{\begin{equation}}
\newcommand\bear{\begin{eqnarray}}
\newcommand\eeq{\end{equation}}
\newcommand\eear{\end{eqnarray}}
\begin{document}

\title{ Long Range Electron Transfer Across
$\pi$-conjugated chain: Role of Electron Correlations}

\author{ S. Lakshmi, Ayan Datta and Swapan K. Pati}
\address{Theoretical Sciences Unit and Chemistry and Physics of Materials Unit,
Jawaharlal Nehru Centre for Advanced Scientific Research,
Jakkur Campus, Bangalore 560 064, India.}

\date{\today}
%\widetext

\begin{abstract}
%\baselineskip 24pt
%\parbox{6in}
{ We consider a prototype polyene chain:
donor-$\pi$(bridge)-acceptor. The distance between the donor and
the acceptor is varied by increasing the number of atoms in the
bridge and the rate of electron transfer, k$_{et}$ is studied for
a series of different donors, D=NH$_2$, OCH$_3$, SCH$_3$, and a
fixed acceptor, A=NO$_2$. We observe a large k$_{et}$ even at a
D-A separation of $\sim$ 45 $\AA$, unexpected from the standard
noninteracting theories. Such a long range electron transfer is
primarily due to the formation of bound electron-hole pair
(exciton) across the donor-bridge-acceptor atoms. Calculations at
various levels, semi-empirical (with inclusion of configuration 
interactions) and many-body models (using density matrix 
renormalization group, DMRG), have been performed to accurately 
account for such electron correlations.}
\end{abstract}
\maketitle
%\narrowtext

\section{Introduction}

 Electron transfer (ET) between well separated donor and acceptor groups is
dominant in many chemical and biological processes. These
processes include design of electronic switches and wires in
molecular level to photosynthesis and respirations in biological
systems\cite{intro,intro1}. Typically in many systems of interest,
the spatial separation between the donor and the acceptor ranges from
10 to 30 $\AA$\cite{distance}. Because of this separation, there
is no direct overlap between the donor and the acceptor groups.
The electron transfer coupling is mediated {\it via} the intervening
orbitals, called the bridge states. In most of the studies, the
model used is a donor-bridge-acceptor (DBA) complex, wherein the
donor and the acceptor states couple to the bridge in the form of
a macromolecule. The intervening bridge can be water, conjugated
organic molecules and polymers, a protein backbone, or even DNA.
In many of the systems, these bridge states can lead to
substantially altered electron transfer rates $K_{et}$ with
respect to vacuum tunnelling\cite{bridge}. The rate is also known
to significantly depend on the conformational degrees of freedom
of the bridges\cite{condon_approx}. However, the exact role of the
bridges in controlling electron transfer is very poorly understood
particularly for systems with strong electron
correlations\cite{corr} and weakly bonded species like the
hydrogen bonded systems\cite{hbond}.

For most systems, these electron transfer rates fall off
exponentially with distance ($\ln {k_{et}}\propto -\beta r$). The
inverse decay length, $\beta$, typically varies from $0.2$ to $1.5$
$\AA^{-1}$, depending on the systems under consideration. Small
$\beta$ values have been observed for many proteins, artificial
peptide chains, DNA segments or a conjugated oligomeric
backbone\cite{isied,longrange}. This exponential decay appears in
the Marcus theory of non-adiabatic electron transfer \cite{Harriman} 
that assumes that the rate of ET depends on the electronic 
coupling element, $H_{DA}$ according to 
\bear
k_{DA}(R_{DA})=\frac{2\pi}{\hbar}|H_{DA}|^2FC 
\eear 
\noindent
where FC is the Franck-Condon-weighted average density of states.
The electronic tunnelling matrix element in turn is related to the
distance of separation between the D and A as $H_{DA}\propto
\exp{-\beta r/2}$. The theory assumes that the time scale of the
electron transfer process is much faster than the time associated
with the reorganization of the solvent nucleus or thermal
disordering.

To date, most theoretical work has focused on the properties of
the bridging medium and the position of D/A levels with respect to
the HOMO and LUMO energies of the bridge\cite{ettheory}.
Furthermore, these theories assume that the role of the bridge is
to provide virtual orbitals that create an effective electronic
coupling between the donor and the
acceptor\cite{marcus,kemp,newton}. This mechanism often reduces
the problem of the whole system consisting of the donor, acceptor
and the bridge subspaces to an effective two-state Hamiltonian
consisting only of the donor and the acceptor states. There have
been some alternative approach to this charge-transfer problem in
recent years. Davison et al. have studied the charge-transfer
thorugh a H\"uckel and extended model bridge using Green's
function formalism\cite{davison}. On the otherhand, a completely
new approach for an extended bridge with narrow band donor and
acceptor groups predict the charge-transfer mechanism to be
long-ranged, involving a Kondo-singlet across the donor (acceptor)-bridge
atoms\cite{myprl}. 
Most often however, in many of the models, explicit
electron-electron interactions in the bridge are completely
ignored, while it is well known that electron correlations often
lead to qualitatively different descriptions.

In this work, we perform a quantitative estimation of the role of
the bridge as well as the donor/ acceptor groups for an effective
charge transfer across the bridge. The bridge energetics has been
found to influence the strength of the donor-acceptor coupling
because of strong mixing between donor/acceptor and bridge
orbitals\cite{padden_row,jortner}. Furthermore, the most
frequently used bridges are the $\pi$-conjugated chains, which are
known to be highly correlated systems\cite{corrln,et_corr}. In the
DBA system, the process of electron transfer takes place with an
electron originally at the donor state, transferring to the
acceptor state {\it via} the bridge states. In the presence of
interactions, the probability of electron transfer to an empty
site is larger than it is to a partially occupied site. The
probability of the relevant site being empty, partially occupied
or doubly occupied depends very strongly on the electron
correlation strength. Thus, incorporating electron-electron
interactions is crucial to the study of electron-transfer
processes.

In the next section, we discuss the electron-transfer process in
a $\pi$-conjugated bridge of varying length with a range of donor
strengths. We find that there is a significant amount of ET for 
$\pi$-conjugated system even when the D and A are separated by a 
very large distance ($\sim$ 45 $\AA$). In 3rd section, we study 
the charge-transfer process for a model conjugated chain with and
without electron-correlations by employing the density 
matrix renormalization group (DMRG) method\cite{white} 
and show that strong electron correlation is indeed required for long-range 
ET. We conclude with a summary of all the results in the 
last section.

\section{Semi-empirical Calculations}

We consider a series of $\pi$-conjugated systems 
shown in Fig. 1. In the top (case 1), the bridge is considered to be a linear
polyene chain of length $n$ varying from 2 to 30. The acceptor
group is fixed as A=NO$_2$, and the donor group is varied,
D=NH$_2$, OCH$_3$ and SCH$_3$. All the geometries are optimized
using the AM1 parameterized Hamiltonian, a part of semi-empirical
MOPAC package\cite{am1}. We have also verified that these
geometries correspond to the minimum energy structure by further optimization
at the B3LYP/6-311G++(d,p) level\cite{becke}. No symmetry
constraints were used in the optimizations. In case 2, we consider
that the donor and the acceptor are separated by a saturated
alkane chain. These two model cases serve as a good contrast in
the mechanism of electron-transfer (ET). The former is expected to
follow the bridge mediated ET due to the extended conjugation
between the donor and the acceptor (through the p$_z$ orbitals of
the intervening atoms). In the following, we show that the 
case 1 is indeed very different and shows long range ET purely because of 
electron correlations.

The geometry optimized structures were used for configuration interaction 
(CI) calculations to
obtain energies and the dipole moments in the CI basis using the
Zerner's INDO method\cite{zerner}. The levels of CI calculations
have been varied but we restrict our discussions to results
obtained using singles CI (SCI) only as this gives a reliable estimate
of the states involved in charge-transfer. The CI approach adopted
here has been extensively used in earlier works, and was found to
provide excitation energies and dipole matrix elements in good
agreement with experiments\cite{mrdci1,mrdci2,mrdci3}. For the
Hartree-Fock determinant, we use varying number of
occupied and unoccupied molecular orbitals to construct the SCI
space till a proper convergence is obtained.

The electron
transfer coupling element, $H_{DA}$, has been calculated using the
generalized Mulliken-Hush method for two-states, referred as 1 and
2\cite{shin}.
\beq
H_{DA}=\frac{\mu_{12}{\Delta
E_{12}}}{\sqrt{{\Delta\mu_{12}}^2+4\mu_{12}^2}}
\eeq
\noindent where
$\Delta\mu_{12}$($\Delta E_{12}$) is the dipole moment (energy)
difference between the two states, and $\mu_{12}$ is the
transition dipole between states 1 and 2. As has been discussed
earlier, the method considers only the adiabatic states 1 and 2,
and assumes that the contribution from the nonadiabatic states to
the transition dipole is small enough since the dipole moment
operator has a weak distance dependence. Note that, it depends on
the calculations whether the nonadiabatic states are well
described by the adiabatic states 1 and 2.

The plots of $H_{DA}$ with respect to the number of CH unit of the
polyacetylene bridge are shown in Fig. 2 for D=NH$_2$, OCH$_3$,
SCH$_3$ and A=NO$_2$. All of them have an exponential decay
profile for the separation between the D and A. Fitting a single
exponential parameter $\beta$, for these three donors, we get
$\beta$=0.277$\AA$$^{-1}$, 0.219$\AA$$^{-1}$ and 0.237$\AA$$^{-1}$ 
for D=NH$_2$, OCH$_3$, SCH$_3$ respectively. Thus, $H_{DA}$ seems 
to depend on the donor strength although for all three cases we 
find the range of electron transfer quite substantial.

A very important inference that can be drawn from our calculation
is that even for a bridge length of 45-50 $\AA$, the $H_{DA}$
magnitude is quite significant ($\sim$ 1.7) and this value is
almost independent of the electronegativity (donor ability) of the
ligands. This was also evident from the very low decay exponential
parameter, $\beta$ $\sim$ 0.25. Such low decay features for ET
have been reported for DNA\cite{bart} and has triggered a lot of
interest in its molecular electronic properties\cite{molelect}. We
propose here that these D and A separated $\pi$ bridge is also a
very good candidate for molecular electronics. While polaron
assisted ET is believed to be the mechanism in DNA, we suggest the
role of electron-electron interactions leading to the formation of
an electron-hole pair (an {\it exciton}\cite{exciton}) responsible for long
range charge transfer in conjugated polymeric systems.

The formation of exciton is evident from our CI calculations. We
have analyzed the oscillator strength of the lowest few CI states for
various system sizes. At smaller chain lengths, the lowest dipole
allowed state (the state that has maximum oscillator strength)
lies quite high in energy and very close to the single-electron
continuum. However, for large chain length (n $>$ 4), the exciton
stabilization is quite large and the dipole allowed state appears
much below the continuum. In fact, as the bridge length increases
from $n=2$, the lowest dipole allowed state shifts its position in
the energy spectrum and above $n=4$ becomes the lowest energy
(fundamental) excitation of the system. For smaller chain length,
the exciton formation is very weak, since such a small length
scale does not provide room for formation of a well bound electron-hole
pair. In fact, a critical length scale is necessary for a
delocalized electron and and a delocalized hole to bind. 
In contrast to this feature
for the conjugated chains (case 1), for the saturated chains
H$_2$N-(CH$_2$CH$_2$)$_{n}$-{NO}$_2$ (case 2), there are no excitonic bands
and the dipole allowed state remains 
invariant of the chain length
enhancements. In fact, the dipole allowed state itself has a small
oscillator strength (very weak). This is easy to understand as for the
saturated chains, the electron and hole over the D and A
respectively, remain uncorrelated for any bridge lengths.

Interestingly, the actual decay of ET in H$_2$N-(CH=CH)$_{15}$-{NO}$_2$ is
much smaller. The $H_{DA}$ at a separation of 1.41 $\AA$ is 3 (From Fig. 2)
and that at 38.5 $\AA$ is 1.7. Using the equation (1) and assuming
that the Franck-Condon rule to be valid, we get the decay to be
only 32$\%$ to that for no separation. This is very different from 
the case 2 in Fig. 1, where the D and A are separated by a saturated 
spacer and the rate decays to zero after n=3. 

Such a feature is very well illustrated in Fig. 3, where we plot the
wavefunctions corresponding to highest occupied molecular orbital
(HOMO) and lowest unoccupied molecular orbital (LUMO) for two of
the model systems: D and A separated by unsaturated and
saturated spacers by large distance:
H$_2$N-(CH=CH)$_{15}$-{NO}$_2$ and
H$_2$N-(CH$_2$CH$_2$)$_{15}$-{NO}$_2$ respectively. The HOMO plot
shows the ground state features with electron localized around the
donor while the LUMO plot shows the redox process, the lowest
charge transfer state. There is a distinct difference between the
two systems. While for H$_2$N-(CH$_2$CH$_2$)$_{15}$-{NO}$_2$,
the electron is localized either on the D or A corresponding to
HOMO and LUMO respectively, for
H$_2$N-(CH$_2$CH$_2$)$_{15}$-{NO}$_2$ shows very large
delocalization of charge carriers from the D and A groups into the
bridge orbitals.

In Fig. 4, we show the molecular orbitals (MO) coefficients for the
atoms in donor and the acceptor groups (N atom in NH$_2$ group and
N and O atoms in NO$_2$ group). The MO coefficients,
$\sum_i{c^{\dag}_{mi}c_{mi}}$ (where $i$ index runs for each
atomic site and $m$ indexes the molecular orbitals), are extracted
for both HOMO and LUMO states. The HOMO state corresponds to the
HF ground state and thus represents the system before charge
transfer. On the otherhand, the corresponding LUMO coefficients
represent the state after charge transfer. As can be seen, the MO
coefficients before charge transfer are mostly located near the
electron donating $\pi$ orbital of N atom in NH$_2$ group. After
the charge transfer however, the electrons are distributed over
the p$_z$ orbitals of three atoms: N and two O of the NO$_2$ group.

\section{Quantum Many Body Calculations}

For a quantitative understanding of the charge-transfer processes
in such systems, we have used quantum many-body models. We
consider the bridge to be a one-dimensional chain with $N$-atoms
($N$ even) and the donor(D) and the acceptor(A) atoms are
attached to its two ends. The Hamiltonian can be written as

\bear
H & = & -t\sum_i[\{\sum_{\sigma}(c^{\dag}_{i\sigma}c_{i+1\sigma}+hc)\} \nonumber \\
& + & Un_{i\uparrow}n_{i\downarrow}+V(n_i-n_{av})(n_{i+1}-n_{av})] \nonumber \\
& + & \epsilon_Dn_D +\epsilon_An_A + \sum_{\sigma}[t_1 (c^{\dag}_{1\sigma}
c_{D\sigma}+c^{\dag}_{N\sigma}c_{A\sigma}+hc)]
\eear

\noindent where $t$ is the transfer integral, $U$ the Hubbard on-site
potential and $V$ is the nearest-neighbor Coulomb interaction parameter
in the bridge. $\epsilon_D$ and $\epsilon_A$ are the on-site energy
of donor and acceptor atoms respectively. $t_1$ is the hopping
term between the $D(A)$ atoms with the bridge atoms $1(N)$.
$c^{\dag}_{i\sigma}$ creates an electron with spin $\sigma$ at site $i$,
and $n_i=n_{i\uparrow}+n_{i\downarrow}$, is the electron
number operator at site $i$.

The $U-V$ model or the extended Hubbard model is the minimal model
to capture the physics of a quasiparticle like exciton, a bound
electron-hole pair. In the large $U$ $(U=\infty)$ limit, the
exciton is stabilized by $V$, since the Hubbard repulsion $U$
describes the electron-hole continuum. However, this is for an
isolated bridge with half-filled band. When the donor(D) and the
acceptor(A) species are very weakly attached to the bridge, the
electron transfer from D (hole transfer from A) can be mediated
through the non-interacting electron-hole states or an excitonic
state of the bridge. However, depending on the electron
correlations in the D and A orbitals, the probability of
intermixing of the D and A states with the bridge states can vary
and to account for the electron transfer in such cases, we need to
consider the whole system as a macromolecule coupled strongly with
its constituents. In our calculations, we have considered two
extreme cases for the donor and acceptors in terms of its electron
correlations. Case A with no Coulomb repulsion at the donor and
acceptor sites which allows any occupancy of the D/A sites and
case B where we have set infinite Coulomb repulsions in D and
A\cite{metalU}, which prevents double occupancy of these sites.
Note that for both cases, bridge is described by correlated model.
The other case with uncorrelated bridge has not been studied,
since it has been described elsewhere.\cite{myprl}

The bridge is assumed to be a half-filled conjugated polymer
insulator. From the half-filled extended model literature, it is
known that the ground state is a SDW (spin-density wave) state for
$U>2V$, while it is a CDW (charge-density wave) state for $U<2V$.
We are interested in charge transfer mechanism through a bridge
with SDW ground state representing organic conjugated polymer, and
thus we set $U=5$, $V=1$ and $t=1$, without loss of
generality\cite{corr}. The total number of electrons in the
D-bridge-A system is $N+1$, where $N$ is the number of orbitals in
the bridge sites and $N+2$ is the total number of orbitals of the
whole DBA system. This system thus represents a linear correlated
chain with one less electron than the half-filled limit. We
present below the charge transfer efficiency of this system with
bridge chain length varying from 2 to 30.

The charge transfer occurs mostly through the lowest energy excited state
of the bridge, which is induced experimentally by interaction
with light. The density matrix renormalization group (DMRG) method
has been used to study the ground and the excited states of
the system\cite{white}. Since the bridge is a one-dimensional chain,
the accuracy
of the method is extremely good. In DMRG method, we target the
ground state and the lowest dipole allowed states, and
the density matrix is averaged over these states at every iteration.
The dipole allowed state is recognized as the state to which the
dipole moment operator acting on the ground state gives nonzero value.
In the DMRG basis the transition dipole moment between two states is
defined as
\beq
\mu_{ij}= \langle \psi_i \mid \hat{\mu} \mid \psi_j\rangle
\eeq

\noindent where $\hat{\mu}$ is the dipole moment operator acting
between the states $\psi_i$ and $\psi_j$. The $H_{DA}$ is then
calculated from the Mulliken-Hush formula given in eqn 2.

We have plotted the electron-transfer coupling element ($H_{DA}$),
defined earlier, as a function of chain length for the two cases
discussed above. Fig 5a shows the $H_{DA}$ for the case with no
Coulomb repulsion at the donor and acceptor sites (case A), for a
few representative values of the donor strengths (orbital energies
$\epsilon_D$). It can be seen that the $H_{DA}$ falls off quite
rapidly as expected, but is quite short ranged. An exponential fit
to the curves give a value of around 0.5 \AA$^{-1}$ for $\beta$,
the parameter in the exponential. This type of large $\beta$
exponential fall of electron transfer rate is expected from the
well-known Marcus' theory\cite{marcus}. Fig 5b shows similar plots
for the case with large Hubbard repulsion at the donor and
acceptor sites (case B). In this case too, the $H_{DA}$ falls off
exponentially, but the fall is much slower and long ranged. An
exponential fit to these curves give a $\beta$ value of around 0.2
\AA$^{-1}$. The value of $\beta$ for case A is larger than that
for case B, indicating that the range of transfer is smaller for
the first case. It is very interesting to note that for the
$\pi$-conjugated bridge considered in the previous section, we
obtained $\beta$ to be of the same order as the case B with
correlated bridge. Thus our model calculations are quite justified
in realizing the ET kinetics for real molecular systems with
strong correlations.

To understand the mechanism behind the charge transfer which
results in long range and short range transfer for case B and case
A respectively, we have plotted the ground state charge densities
in Fig. 6 for both cases. Fig 6a shows the ground state charge densities for case
A. It can be seen that the presence of a strong donor (a large
negative value of $\epsilon_D$) and the absence of any Coulomb
repulsion at the donor favors an increase in the charge density at
the donor. The acceptor on the other hand, which is always kept at
the same strength (with a value of $\epsilon_A=0$ as for the
bridge), prefers to retain the average charge density that a
system with one less electron would have, as a change in the
electron number at this site, will not result in any change in the
total energy. The extra charge density at the donor site is
compensated by the reduction in the charge density at the bridge
site next to the donor. The hopping energy between the donor and
the first site of the bridge favors the presence of a hole
(reduced charge density) at the first bridge sites, so that the
system can be stabilized by this kinetic energy. A further
stabilization is effected by the presence of the nearest neighbor
Coulomb repulsion between the donor and this first bridge site. On
the other hand, the last bridge site shows an increased charge
density and the last but one bridge site has a reduced charge
density. This can be attributed to the fact that hopping between
the last bridge site and the acceptor site gets renormalized
because of the absence of any Coulomb repulsion at the acceptor,
and hence stabilizes the system. The reduced charge density at the
last but one bridge site also allows for hopping between this site
and the last bridge site. The presence of the extended Hubbard
Coulomb repulsion energy $V$ between these sites further enhances
the kinetic stabilization. This excess electron and hole is formed
due to the correlations in the bridge; they are not free but are
rather bound. We refer to this kind of a quasiparticle as
``exciton''. The exciton is formed at the ends of the bridge closest
to the donor and the acceptor sites and is not spread over the chain.
It has rather small volume and is quite localized.

On the otherhand, for the case with large Coulomb repulsion at the
donor and acceptor sites (case B), the ground state charge
densities show contrasting behavior, as shown in Fig.6b. Here, the
charge densities show a dip (reduction) at both the first and the last
bridge sites closest to the donor and the acceptor. In this case, the
large Hubbard $U$ ensures that there is only a maximum of one
electron which is allowed at the donor and the acceptor sites. No
matter however large the strength of the donor, the charge at the
donor site does not exceed $1$. When the donor and the acceptor energies
are kept the same as the bridge ($\epsilon_D=\epsilon_B=\epsilon_A=0$), 
the charge densities at all
sites, except the bridge sites closest to the donor and the
acceptor, have an average number of electrons that a system with
one less electron from the half-filling would have. This reduction
in charge at the first and last bridge sites is due to the
resulting stabilization that would be brought about by the
presence of the hopping energy $t$ and the extended Hubbard energy
$V$. A higher donor strength shows a similar feature at the first
and the last sites of the bridge, however, a very prominent increase in the
charge density can be seen in the first half of the bridge and a
corresponding reduction in the other half of the bridge. This
shows that the exciton in this case spreads over the size of
almost the length of the entire bridge. This large volume exciton
is in fact responsible for the long range electron transfer for
the case B.

We have also looked at the excited state charge densities for both
cases. For case A, there is not much difference between the ground
and the excited state charge densities indicating that the charge
transfer has not been mediated by the localized exciton, whereas
for case B, the excited state charge densities clearly show the
movement of charge density from the donor to the acceptor with a
reverse in the density pattern from that found for the ground state.

We also have studied the charge-charge correlations functions
between the donor/acceptor and the bridge sites: $\langle n_{D(A)}
n_B \rangle - \langle n_{D(A)} \rangle \langle n_B \rangle$, where
$n_{D(A)}$ and $n_B$ are the charge densities at the
donor (acceptor) and the bridge sites respectively. The ground
state charge-charge correlations in case A for all strengths of
the donor, clearly indicates that the correlations between the
donor and the bridge sites are very short ranged and that the
exciton is not effective in mediating the charge transfer between the far
separated donor and the acceptor. In contrast to this, the
correlations in case B shows features in correspondence with the
charge density in Fig.6b, i.e., the donor is highly correlated
with the bridge sites and hence transfer mediated by the exciton
is the most likely mechanism of charge transfer between the donor
and the acceptor.

Finally, to understand the nature of the excitonic state
responsible for longer range electron transfer, we have plotted
the chain axis component of the transition dipole moment between
the ground state and the first and second excited states for the
two cases discussed above, for a particular donor strength
($\epsilon_D=-4.0$) in Fig.7a and 7b respectively. This transition
dipole moment enters into the calculation of $H_{DA}$ and hence is
a good measure of the strength of the states carrying the charge
for transferring to/from the acceptor end. When there is no Coulomb
repulsion at the donor and the acceptor sites (case A), the
transition dipole moment between the ground state and the first
excited state increases and then falls off for longer chain
lengths, whereas that between the ground state and second excited state
shows a steady increase, clearly indicating that the state
carrying the charge is not a well defined electron-hole bound pair
state. On the other hand, for the case with large Coulomb
repulsion at the donor and the acceptor sites (case B), the
transition dipole moment between the ground and the first excited
state shows an increase with the increase in the chain length,
indicating that this is the state which carries the charge
through, from one end to the other even for large bridge length.
The same between the ground state and the second excited state has
zero value for small system sizes and a very negligible value for
larger chain lengths, clearly showing that this state does not
play any part in the charge transfer between the donor and the
acceptor states. This emphatically proves that for case B, the
charge is carried by a well defined fundamental quasiparticle for
all the chain length resulting in long-ranged electron-transfer.

\section{Conclusions}
To conclude, we have performed semi-empirical as well as many-body
DMRG calculations to understand the role of electron-electron
interactions in governing the electron transfer in D-$\pi$-A
systems. The electron transfer kinetics are truely long-ranged for
strongly correlated conjugated polymeric systems and are not
affected much by the strength of the donors. What matters is the
extent of interactions in the bridge. It is the mediation of a
stable electron-hole pair quasiparticle (exciton) in these systems
that is responsible for small inverse decay length, $\beta$. We
have been able to quantify the nature of the exciton from
different levels of calculations. The CI calculations clearly
indicate the formation of the excitonic state for an optimum size
of the conjugated bridge length. But for the saturated bridge
case, electrons are localized at the respective sites and thus the
electron transfer has a much shorter range. From the quantum
many-body long-chain DMRG calculations on systems with small and
large correlations between the donor (acceptor) and the bridge
sites, we have been able to distinguish the nature of the quasiparticle
responsible for the electron transfer in the two cases. Our
results show that when the donor (acceptor) and bridge are weakly
correlated, the charge-transfer is purely by uncorrelated electron
and hole, leading to short range ET observed in earlier theories. On
the other hand, when the donor (acceptor) and bridge sites are
strongly correlated, formation of well-bound electron-hole pair
spread over the entire chain length is involved in the charge
transfer, even for larger chain lengths, giving a very long-range
ET. Thus, the exciton formation is a crucial parameter in
controlling long-range ET. Formation of such a quasi-particle is
known to lead to very interesting optical properties in organic
light emitting diodes. The fact that excitons also play an
important role in transport mechanism signifies an important
contribution towards fabrication of molecular wires where currents
can be induced by optical excitations. Also, for biomolecular
systems with known long-range ET like DNA, the identification of
right kind of quasiparticle responsible for charge and field
carrying processes can lead to novel applications in bio-molecular
optoelectronics. This however, requires further study.

\pagebreak
\clearpage

\begin{figure}
\includegraphics[scale=0.6,angle=0] {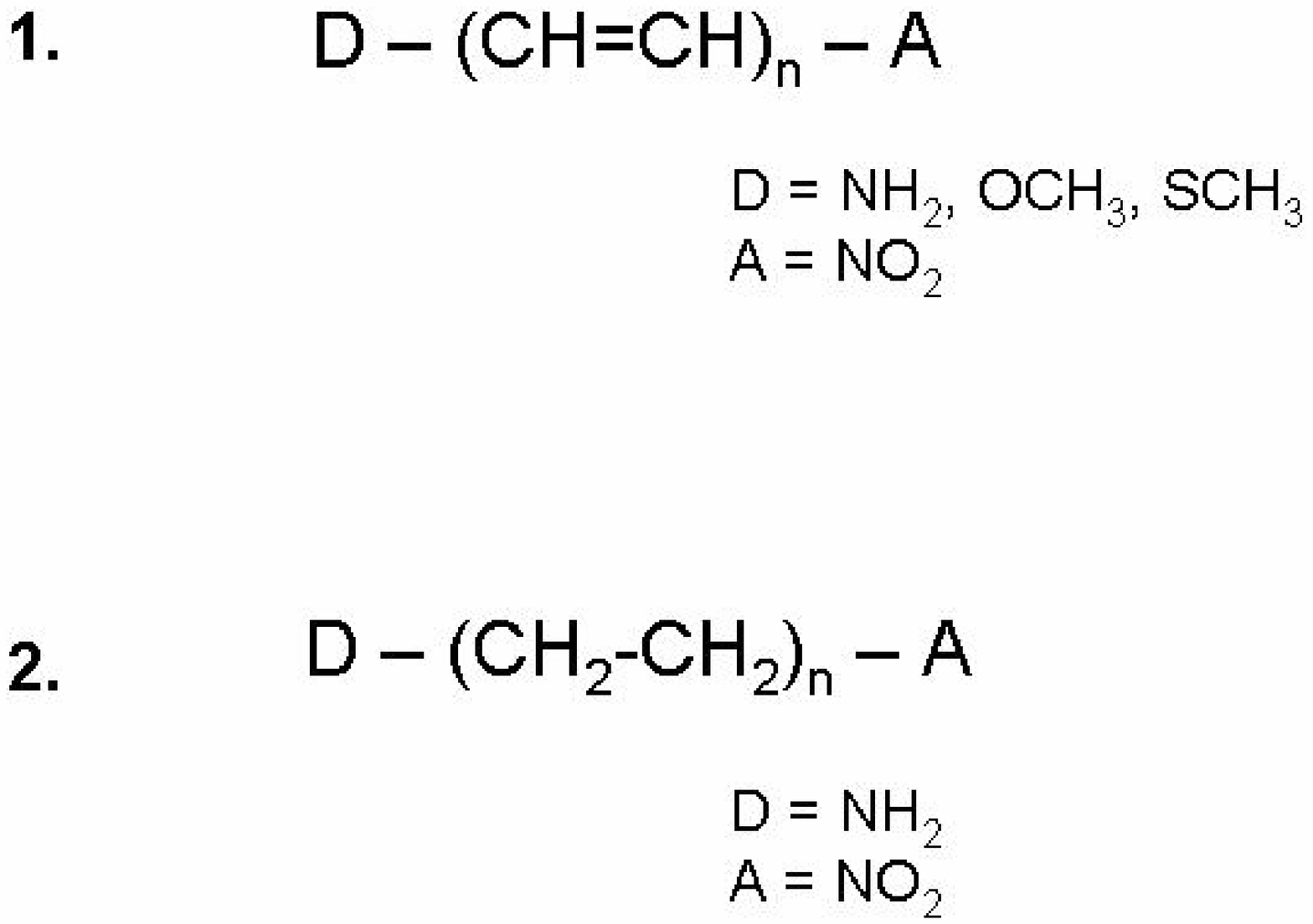}

\caption{Model systems considered for study}

\end{figure}

\newpage
\clearpage

\begin{figure}
\includegraphics[scale=0.6,angle=270] {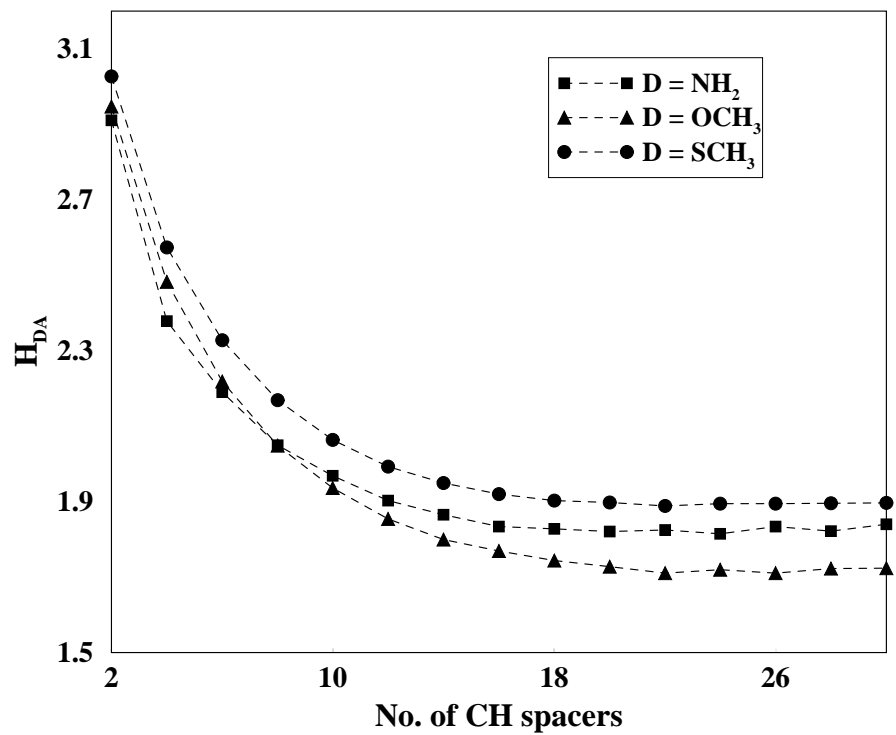}

\caption{ET coupling matrix, H$_{DA}$ (in eV) with respect to the increase in
the number of olefinic spacers for different donors, D and fixed A=NO$_2$ }

\end{figure}

\newpage
\clearpage

\begin{figure}
\includegraphics[scale=0.6,angle=0] {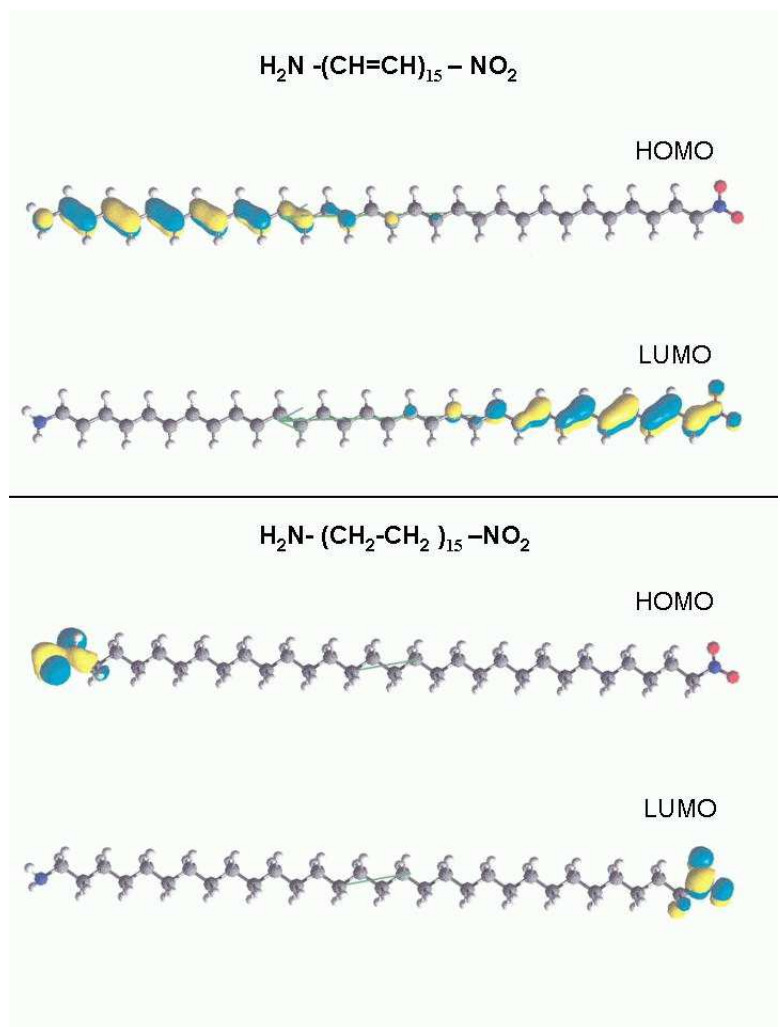}

\caption{Frontier orbital plots with unsaturated and saturated spacers. The
shading of the lobes show the parity. Atom symbol: Red: Oxygen, Blue: Nitrogen,
Black: Carbon, White: Hydrogen. Light green arrow shows the direction
of the dipole moment}

\end{figure}

\newpage
\clearpage

\begin{figure}
\includegraphics[scale=0.6,angle=270] {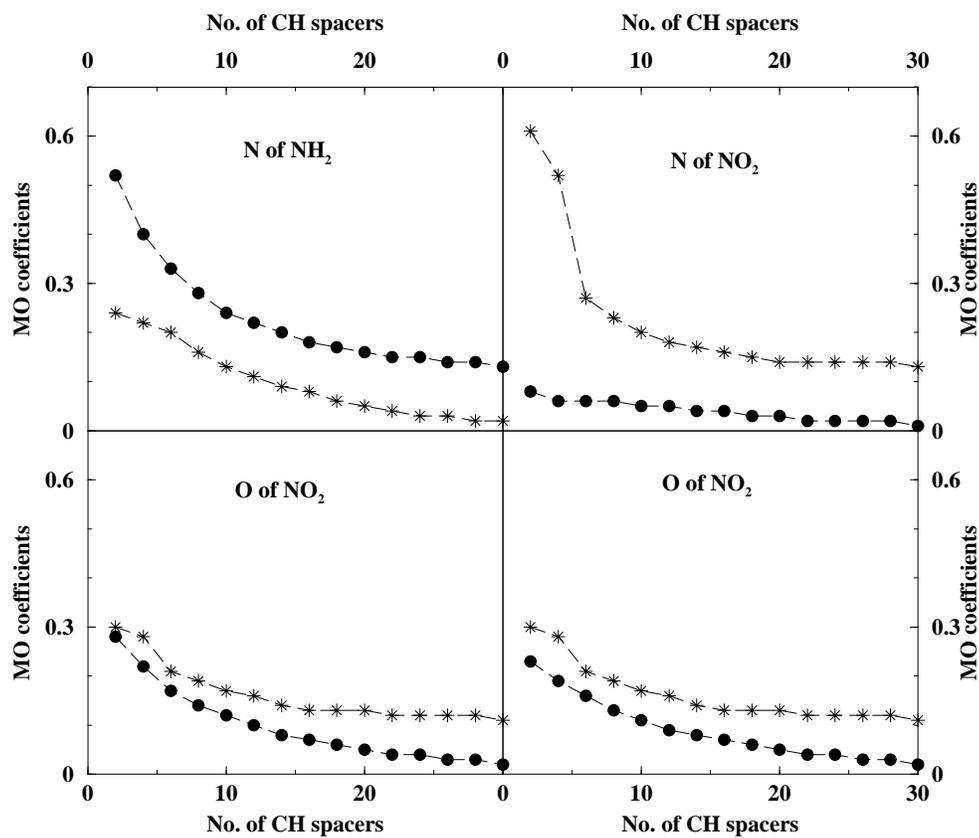}

\caption{Coefficients of molecular orbital occupancy on the N atom of NH$_2$
and the N atom and two O atom of NO$_2$ before (circles) and after
(stars) ET.}

\end{figure}

\newpage
\clearpage

\begin{figure}
\includegraphics[scale=0.6] {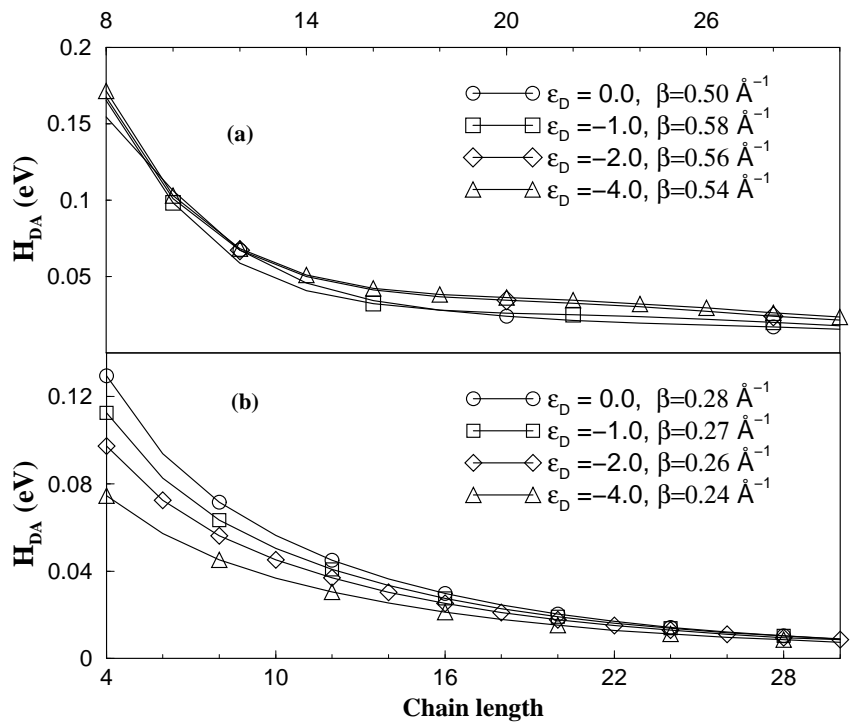}

\caption{ ET coupling matrix $H_{DA}$ with respect to the bridge length
for case 1 (above) and case 2 (below) for different values of the
donor strengths ($\epsilon_D=0.0$ to $\epsilon_D=-4.0$)}
\end{figure}

\newpage
\clearpage

\begin{figure}
\includegraphics[scale=0.6,angle=270] {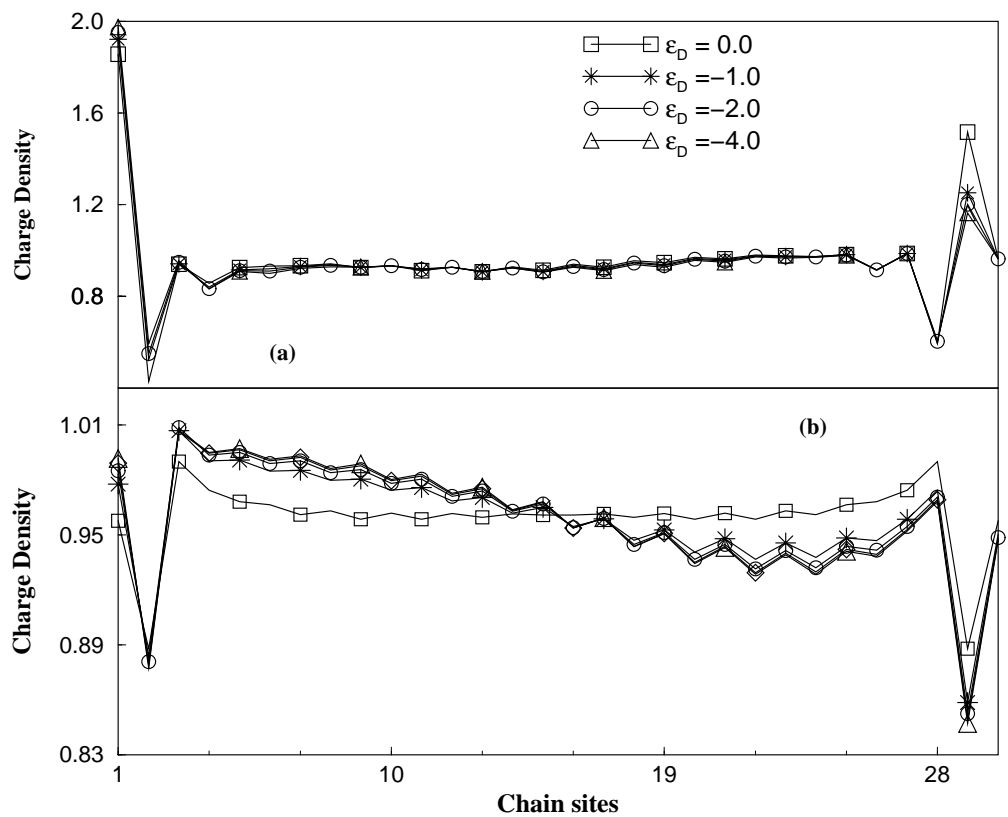}

\caption{ Charge densities at every bridge site
for case 1 (above) and case 2 (below) for different values of the
donor strengths ($\epsilon_D=0.0$ to $\epsilon_D=-4.0$)}

\end{figure}

\newpage
\clearpage

\begin{figure}
\includegraphics[scale=0.6,angle=270] {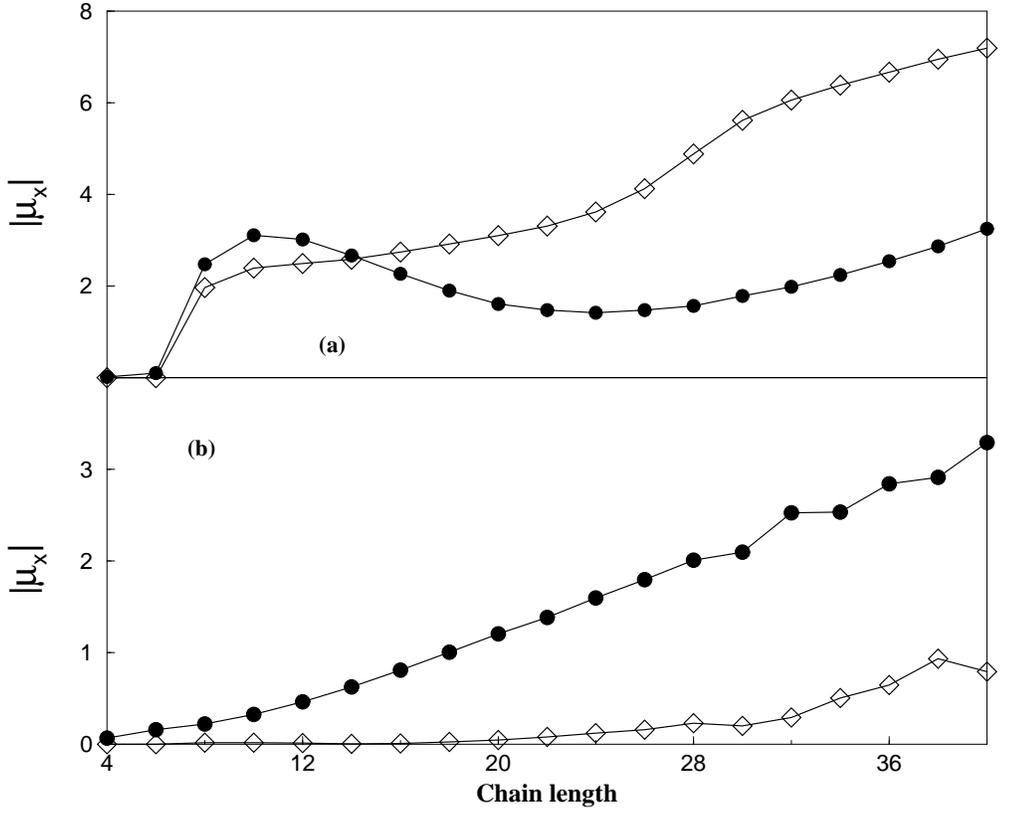}

\caption{ Modulus of the x component of the transition dipole moment,
$|\mu_x|$ with respect to chain length for case 1 (above) and case 2 (below)
for transition from ground state to first excited state (filled circles) and ground
state to second excited state (diamonds)}

\end{figure}

\newpage
\clearpage

\end{document}